\documentclass[%
preprint
 ,secnumarabic%
,amssymb, amsmath,nobibnotes, aps, prb]{revtex4}

       \usepackage[pdftex]{graphicx}

\usepackage{bm}%
\expandafter\ifx\csname package@font\endcsname\relax\else
 \expandafter\expandafter
 \expandafter\usepackage
 \expandafter\expandafter
 \expandafter{\csname package@font\endcsname}%
\fi

\begin{document}

\title{Ripples and Charge Puddles in Graphene on a Metallic Substrate}
\author{S. C. Martin$^1$, S. Samaddar$^1$, B. Sac\'ep\'e$^1$, A. Kimouche$^1$, J. Coraux$^1$, F. Fuchs$^2$, B. Gr\'evin$^2$, H. Courtois$^1$, and C. B. Winkelmann$^1$}%
\affiliation{$^1$ Universit\'e Grenoble Alpes, Institut NEEL, F-38042 Grenoble, France}
\affiliation{$^2$ UMR 5819 (CEA-CNRS-UJF) INAC/SPrAM, CEA-Grenoble, 38054 Grenoble Cedex 9, France}
\date{\today}%
\begin{abstract}

Graphene on a dielectric substrate exhibits spatial doping
inhomogeneities, forming electron-hole puddles. Understanding and controlling the latter is of
crucial importance for unraveling many of graphene's fundamental properties at the Dirac point.
Here we show the coexistence and correlation of charge puddles and topographic ripples in graphene decoupled from the metallic 
substrate it was grown on. The analysis of interferences of Dirac fermion-like electrons yields a linear dispersion relation, indicating that graphene on a metal can recover its intrinsic electronic properties.

\end{abstract}

\maketitle

The study of electron-hole puddles in graphene \cite{Martin2007} has so far relied on graphene sheets isolated by
mechanical exfoliation of graphite on dielectric substrates such as SiO$_2$. The origin of these has been subject to debate, as different studies have pointed to either charged impurities between graphene and SiO$_2$  as sources of the puddles \cite{DasSarma2011,Zhang2009}, others invoking in addition the mixing
of the $\pi$ and $\sigma$ orbitals due to local curvature \cite{Deshpande2009, Gazit2009, Gibertini2010,Gibertini2012, Kim2008, Partovi-Azar2011}. 
In this context, the limited knowledge about the graphene/SiO$_2$ interface and the ensuing low graphene mobility calls for the use of other substrates. 
More generally, experiments based on different dielectric environments, that is, different strengths of charged impurities' screening, have been performed. They however showed no significant influence of the substrate dielectric constant on the graphene electronic properties, thereby questioning the role of charged impurities for the puddle formation \cite{Ponomarenko2009,Couto2011} (see also \cite{DasSarma2012}).

On metallic supports, the origin of charge disorder might be very different. Periodic ripples, arising from the lattice parameter mismatch between graphene and most transition metal surfaces,
were for instance related to the puddles \cite{VasquezdeParga2008} in graphene on Ru(0001). It however turned out that
charge carriers  in this system do not exhibit Dirac fermion-like properties, due to a strong hybridization
between the $4d$ Ru and $p_z$ C orbitals \cite{Wang2008,Sutter2009}. Even in less strongly coupled systems, interaction between
graphene's conduction/valence bands with surface states of the metal \cite{Pletikosic2009,Varykhalov2012} cannot be excluded. 
A linear dispersion relation in the electronic band structure at the
Brillouin zone corners
was recovered in graphene on Ru(0001) intercalated with an atomic layer of oxygen below the graphene \cite{Sutter2010}. This layer unfortunately suppresses the graphene ripples, which prevents from addressing the possible relationship between
puddles and topography. 

In this letter we report on a STM/STS study of corrugated graphene lying on an Ir metallic substrate. The analysis of the quasi-particle interference pattern reveals the linear dispersion relation of the graphene band structure, and demonstrates the absence of hybridization with the Ir substrate. Despite the immediate proximity of the metal that acts as an electrostatic screening plate, we observe electron-hole puddles close to the charge neutrality point, akin to those observed on graphene on dielectric substrates.  Furthermore, the topographic images of the ripples exhibit strong spatial correlations with charge density inhomogeneities, suggesting electron-hole puddles of a new origin.

\begin{figure}[t]
  \includegraphics[width=\columnwidth]{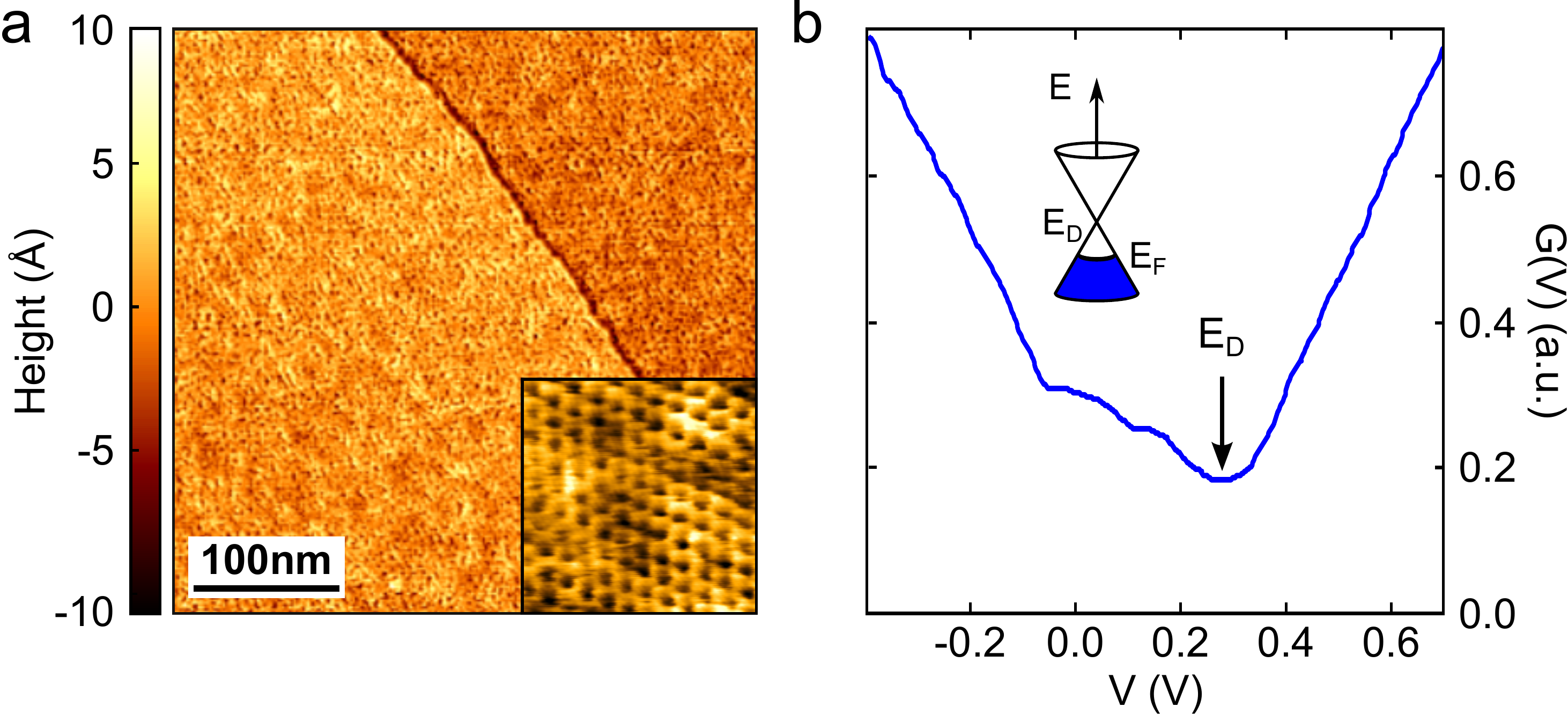}
 \caption{{\bf Topography and density of states of decoupled graphene on iridium.} (a) Scanning tunneling micrograph on graphene on
Ir(111). The
image size is 400$\times$400 nm$^2$, tunnel current $I=1$ nA, bias voltage $V=0.57$ V). The atomic step in epitaxial Ir(111)
covered by graphene is
2.5 \AA \, high. Inset: zoom-in (2.3$\times$2.3 nm$^2$) atomic resolution image, $I=1$ nA, $V$=0.01V.  (b) Local tunneling spectroscopy $G(V)=dI/dV$. The Dirac point ($eV=E_D$, arrow) is defined by the minimum of $G$.}
 \label{topo}
\end{figure}

The studied sample  is graphene prepared by chemical vapor deposition on epitaxial Ir(111) under ultra-high vacuum (UHV), as described in \cite{Vo-Van2011}.
Exposure to air can decouple the graphene
from its metallic support by oxygen intercalation, transforming the moir\'e pattern into a disordered topographic landscape \cite{Kimouche2014} (see also Supp. Info.). 
Figure \ref{topo}a shows the topography of graphene on a plain Ir atomic terrace. A disordered topographic landscape with a typical rms
roughness $\sigma_z\approx50 - 100$ pm is observed, comparable to the corrugation of graphene on SiO$_2$. 
While  the usual moir\'e pattern is absent, atomic resolution is routinely achieved, demonstrating the cleanliness of the surface.

The electronic density of states  can then be accessed in STM by the measurement of the local tunneling conductance $G(V)=dI/dV$.
The tunnel spectra display a V-shape  (Fig.  \ref{topo}b), characteristic of the
density of states of graphene and similar to data observed by STS on exfoliated graphene on SiO$_2$
\cite{Jung2011, Luican2011}. The voltage $V_{min}$ at the conductance minimum has been identified in previous works \cite{Zhang2009} as the charge neutrality point, that is, the Dirac point in graphene at energy $E_D=eV_{min}$. The value of $E_D$ reflects the
strength of the
local graphene doping \cite{Zhang2009,Luican2011,Deshpande2009}, which is $p$-type here. 
A dip of width of about $20$ meV is also present around zero bias, which is frequently reported in STS studies on graphene \cite{Jung2011,
Zhang2008, Luican2011} and will not be discussed in this work.
We have measured a series of high-resolution conductance maps at low temperature.
At each position $\bf r$ of a tunnel conductance map, the Dirac point energy $E_D$ is extracted
using a parabolic fit around the minimum of $G$. The spatially averaged doping ${E_D^0}$ = 340 meV is larger than inferred from
ARPES studies in UHV, which reported $E_D$
at 100 meV for graphene on Ir(111) \cite{Pletikosic2009}. 

\begin{figure}[t]
  \includegraphics[width=\columnwidth]{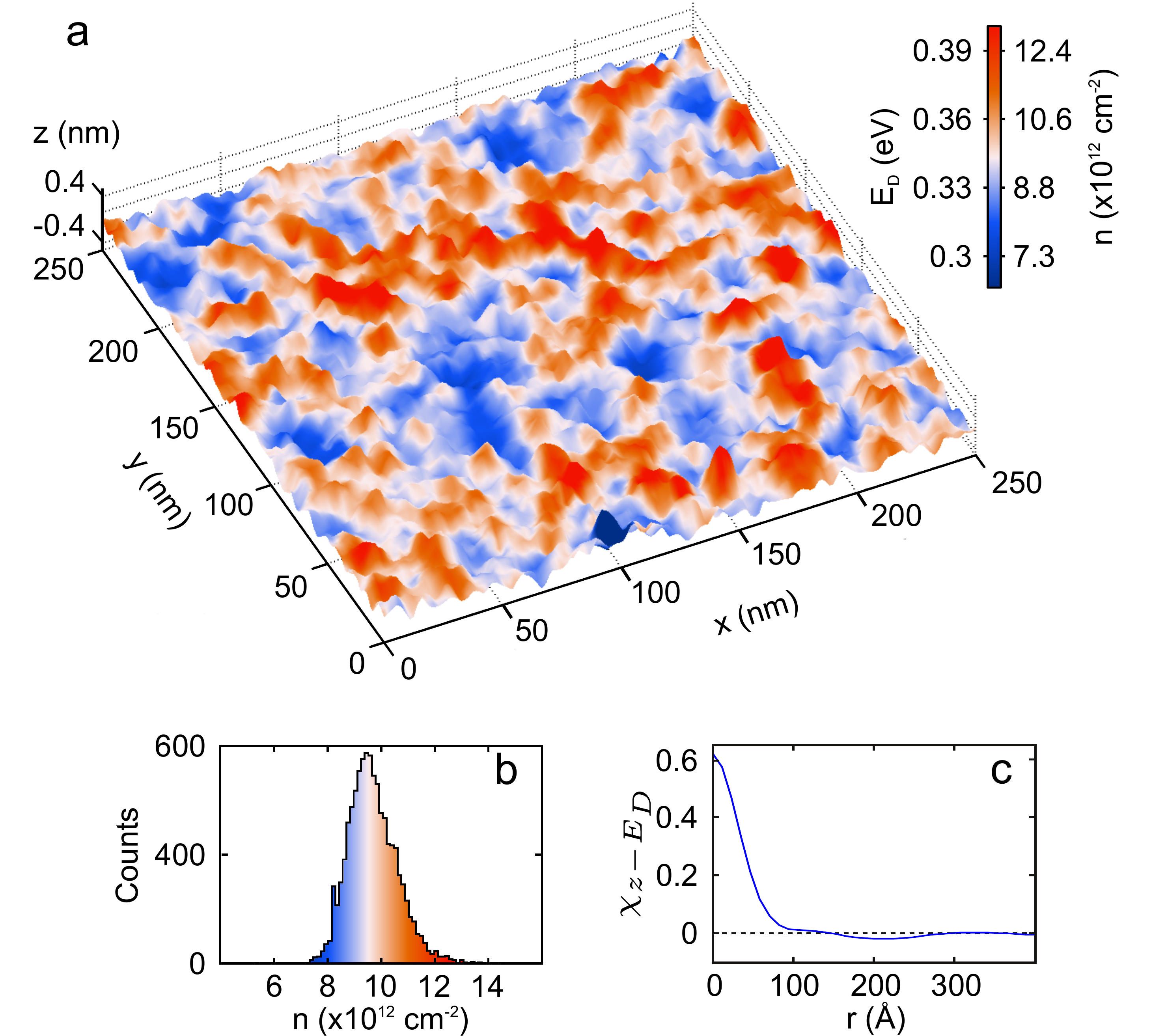}
 \caption{{\bf Spatial inhomogeneities of the Dirac point.} (a) Dirac point map (color code) superimposed with a 3D plot of the
long-wavelength topography. Image of 250$\times$250 nm$ ^2$. (b) Carrier density distribution extracted from (a) (see text).
(c) Angular average $\chi_{z-E_D}(|{\bf r}|)$ of the cross-correlation function (see text) between the above $E_D$ and topography maps.}
 \label{dirac}
\end{figure}

The map shown in Fig. \ref{dirac}a pictures the spatial inhomogeneities of
$E_D$ around its mean value, forming a smooth landscape of charge puddles, of  about $9$ nm size. From 
$E_D^2 = (\hbar v_F)^2 \pi n$ \cite{DasSarma2011},
where $\hbar$ and $v_F$ are the reduced Planck constant and the Fermi velocity respectively, we extract the distribution of the charge
carrier density $n$ (Fig.
\ref{dirac}b),  where we take $v_F=0.89\times 10^{6}$ m/s as will be discussed later. Notably, the standard deviation
$\sigma_n\approx1.1\times 10^{12}$ cm$^{-2}$ is
slightly higher than in graphene exfoliated on SiO$_2$ ($\approx 4\times 10^{11}$ cm$^{-2}$) \cite{Zhang2009,Tan2007}.

We now focus on the comparison between the Dirac point distribution and the
topography. Since only topographic variations at length scales similar or larger than the typical puddle size can correlate with the charge
inhomogeneities, we filter out  the topographic maps from structures of dimensions below half the mean puddle size.
Figure \ref{dirac}a shows the superposition of an $E_D({\bf r})$ map (color scale) along with the long wavelength-pass filtered 
topography $z({\bf r})$ recorded at the same position (3D profile). A very high degree of correlation between doping and topography is readily seen. We have quantified this
by calculating the normalized
cross-correlation function $\chi_{z-E_D}({\bf r})=\sum_{i}E_D({\bf r_i}-{\bf r})\times z({\bf r})/ (\sigma_{E_D}\times \sigma_{z})$ of the two data sets.
The local cross-correlations $\chi^0_{z-E_D}$ between  $z({\bf r})$ and $E_D({\bf r})$ are in excess of 60 \% in large area maps (Fig.
\ref{dirac}c).
These correlations are independent on the region chosen, but are enhanced
in maps with dimensions much larger than the typical puddle size. 
When correlating spectroscopic maps with topography, one also has to recall that in constant-current STM mode a local DOS variation will
lead
to a change in the tip-sample distance $z$. This can misleadingly induce phantom topographic features. We have carefully analyzed this contribution to be modest however (see Supp. Info. for details).

The above analysis therefore leads to the central result of this Letter: disordered graphene on a metallic substrate displays a strong local correlation between doping and topography.
Several scenarios can be considered for the above correlation. A contribution of curvature effects \cite{Gazit2009, Gibertini2010,Gibertini2012,Kim2008} could for example be anticipated. In the present system, the theoretically expected contribution of this effect is however nearly two orders of magnitude below the observed amplitude of the doping disorder \cite{Polini}. 
The graphene doping could further be due to graphene-metal distance dependent charge transfer from the metallic substrate due to finite electronic wave function overlap. 
Calculations of this effect \cite{Giovannetti2008} yield a qualitatively correct description, including the correct sign of $\chi^0_{z-E_D}$ for graphene on iridium. Scanning Probe measurements at the boundary between coupled and uncoupled graphene on iridium have however shown the graphene to iridium distance to increase by about 1 nm upon decoupling \cite{Kimouche2014}, a distance at which the above scenario would have negligible contributions. Yet another possible scenario might call for doping induced by  molecular species intercalated between the graphene and its substrate \cite{Kimouche2014}. Locally enhanced accumulation of negatively charged intercalates can induce a reduction of the Fermi level, that is, enhanced $p$-doping in the graphene.

In order to obtain a more microscopic understanding of the role of the local electrostatic environment as electron scattering centers,
we have tracked the wave vector $\bf k$ distribution of
interference patterns in the Fourier transforms of DOS maps. The evolution of the dominant $\bf k$, relative to the Brillouin zone corner as a function of the energy level of each map then reflects the
dispersion relation of the scattered electrons. In graphene, both the large $k$ intervalley and the small $k$ intravalley scatterings have been observed by this technique  \cite{Rutter2007}. For intravalley scattering, the wave vector transfer $\bf q$ links two points of the circle resulting from the intersection of a given Dirac cone and a constant energy plane. A particular characteristic of graphene is the suppression of low-energy backscattering \cite{Katsnelson2006}. One therefore expects a smooth distribution of $q$ between 0 and $2k_F$ for intravalley scattering, seen as a disk in reciprocal space.

\begin{figure}[t]
   \includegraphics[width=\columnwidth]{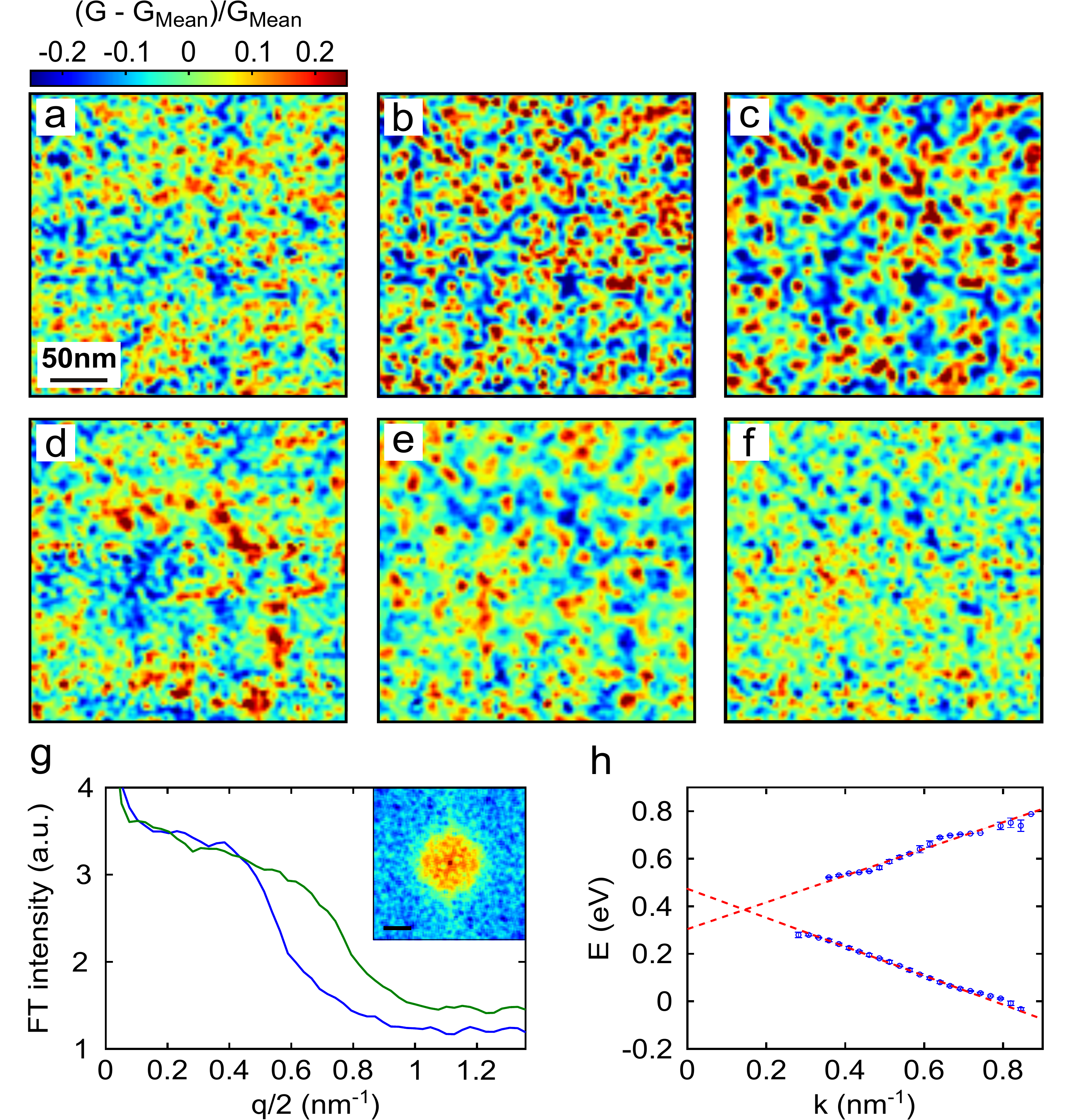}
\caption{{\bf Interferences and dispersion relation.}(a) - (f) $G({\bf r},
E)$ maps for
$E=-185; 15; 125; 345; 555$ and $700$ meV respectively. The color scale is set to cover fluctuations of $\pm25$ \% with respect to the
average $G$ at a given $E$. As the Dirac point is approached, the islands size can no longer be resolved. (g) Power spectral density from angular averaging of Fourier transform of (b) (blue) and (c) (green). The inset shows the Fourier transform of (b). The scale bar is 0.5 nm$^{-1}$. (h) Energy - wave vector
 relation extracted from Fourier analysis of the density of states maps at energy $E$. The dashed lines are fits to the linear dispersion relation of graphene, yielding $v_F=0.89\times 10^{6}$ m/s and $E_D^0=0.38$ eV.}
 \label{DOS-fluc}
\end{figure}

In our data, the DOS maps at energies far from $E_D^0$ display clearly resolved structures at length scales
smaller than the typical puddle size (Figs. \ref{DOS-fluc}a--f). Further, the size of the observed features decreases with increasing
$|E-E_D^0|$. The Fourier transform maps display a disk-like structure (Fig.  \ref{DOS-fluc}g) from which we extract $k=q_{max}/2$ at each energy (Fig.  \ref{DOS-fluc}h). The radius $q_{max}/2$  of the interference patterns is defined as the inflection point of the angular averaged Fourier intensity, that is, the minimum of its smoothed
derivative for 0.25 nm$^{-1}<k<1$ nm$^{-1}$. This criterion produces the $k(E)$ dispersion with least noise but also slightly overestimates the wave vectors involved by adding a constant shift to the detected values of $q$, which results in the linear dispersion bands crossing at $k>0$ in Fig. \ref{DOS-fluc}h.

The experimental dispersion relation at small $k$ has the features of graphene close to the Dirac point, evidencing thus the scattering mechanism at work as intravalley. 
It is actually notable that this can be observed in graphene on a metallic substrate at all. 
A fit by the linear dispersion relation of graphene $E= E_D^0 \pm
\hbar v_F k$ (Fig.  \ref{DOS-fluc}h) yields $v_F=0.89\pm0.04\times 10^6$ m/s. 
In unhybridized graphene, $v_F$ is a sensitive probe of the strength of electron-electron interactions \cite{Castro-Neto2009,DasSarma2011}, which can be screened by a large dielectric constant $\epsilon$  environment.  While $v_F=2.5\times10^6$ m/s has been reported on a low-$\epsilon$ quartz substrate \cite{Hwang2012}, it decreases to about $1.1-1.4\times10^6$ m/s on SiO$_2$  \cite{Zhang2009, Jung2011}, with a limiting value of $0.85\times10^6$ m/s expected for infinite screening \cite{Hwang2012}.
The present system can actually be envisaged as graphene on a dielectric substrate with divergent $\epsilon$, efficiently screening electron-electron interactions.

With the above analysis it is seen that (i) the linear dispersion relation of graphene decoupled from a metallic substrate survives at both positive and negative energies and (ii) electron-electron interactions in the graphene sheet are strongly screened. 
This confirms that the present system behaves much more like graphene on a dielectric substrate rather than a graphene-metal hybridized electronic system, but within a screening environment. Finally, the above observation of unhybridized electronic states in the graphene once more favors the scenario of the doping inhomogeneities due to intercalated species over that of charge transfer from the substrate \cite{Giovannetti2008} - the later assuming a finite wave function overlap between graphene and substrate electrons.

\begin{figure}[t]
   \vspace{-2cm}
   \includegraphics[width=\columnwidth]{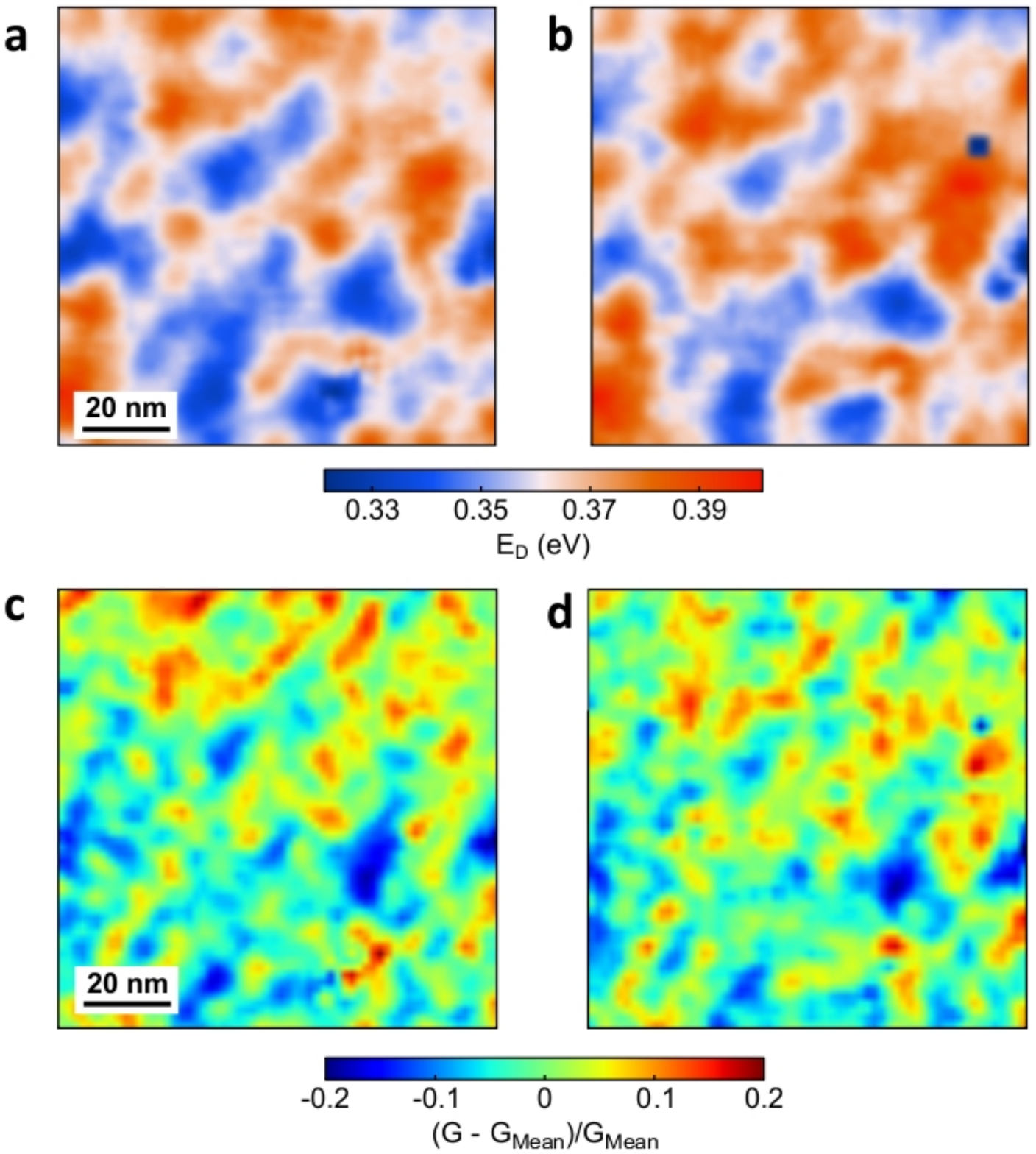}
   \vspace{-2.3cm}
\caption{{\bf Effect of magnetic field.} Dirac point map at zero magnetic field (a) and at $B=2$ T (b). Density of states map at $E-E_D=-330$ meV at zero magnetic field (c) and $B=2$ T (d). All four maps are taken over the same region at $T=130$ mK.}
 \label{field}
\end{figure}

One remaining open question concerns the physical origin of the observed quasi-particle scattering. It is tempting to question localization-type interpretations to explain the DOS features in real space in Fig. \ref{DOS-fluc}. We have studied the variations of the features studied above (puddles and interference patterns) as a function of magnetic field and found no significant change of neither the $E_D$ nor the $G$ maps up to the maximum experimentally accessible magnetic field of 2 T (Fig. \ref{field}). For symmetry reasons, Anderson localization is indeed not expected in graphene in the absence of atomically sharp defects providing intervalley coupling \cite{DasSarma2011}.
The magnetic field-independence of the observed patterns therefore completes the picture of long-range Coulomb interactions generating quasi-particle interferences through intravalley scattering.

In conclusion, the present work demonstrates that CVD grown graphene can be completely decoupled from its metallic substrate, displaying properties in striking analogy with graphene on a dielectric substrate like SiO$_2$.
The main contrasting feature with respect to the latter system is that the dopants here indeed have a topographic signature, which is strongly correlated to the puddle landscape. The linear dispersion relation of graphene indicates perfectly unhybridized graphene with an efficient screening of electron-electron interactions. While the coupling constant of graphene, that is, the ratio of interaction to kinetic energy, $\alpha \propto (\epsilon v_F)^{-1}$ is of order unity in unscreened graphene \cite{Castro-Neto2009}, studying graphene in a large dielectric constant environment could allow for fine tuning of $\alpha$.

This work was supported by the ANR-2010-BLAN-1019-NMGEM and EU-NMP3-SL-2010-246073 GRENADA contracts. The authors are indebted to A. Tomadin, M. Gibertini, F. Guinea and M. Polini for their estimation of the influence of curvature effects. We further would like to thank  L. Magaud, P. Mallet, J.-Y. Veuillen, V. Guisset, V. Bouchiat and P. David  for help and stimulating discussions.

\bigskip

{\bf Methods}

The experimental set-up is a home-made scanning probe microscope operating at very low temperatures (130 mK). Local tunnel conductance data were obtained using the lock-in technique ($V_{AC}=6$ mV, $f=407$ Hz).
Prior to cool-down, the sample is heated to 70$^\circ$C overnight while pumping the chamber. During cool-down, the sample is continuously heated well above all other inner parts of the cryostat in order to avoid cryosorption of residual gases on the sample.

\bigskip

The authors declare no competing financial interests.

\end{document}